\documentclass{XrU2005}
\usepackage{psfig}

\title{The nature of X-ray absorbed starburst QSOs and the QSO evolutionary scheme}
\author{M.J. Page}
\affil{UCL, Mullard Space Science Laboratory, Holmbury St. Mary, Dorking,
  Surrey RH5 6NT, UK}
\author{F.J. Carrera}
\author{J. Ebrero}
\affil{Instituto de F\'isica de Cantabria, Avenida de los Castros, 39005
  Santander, Spain}
\author{J.A. Stevens}
\affil{Centre for Astrophysics Research, University of Hertfordshire, College
  Lane, Hatfield AL10 9AB, UK}
\author{R.J. Ivison}
\affil{UK Astronomy Technology Centre, Royal Observatory, Blackford Hill,
  Edinburgh, EH9 3HJ, UK}

\begin{document}

\keywords{X-rays; submillimetre; galaxies: active}

\maketitle

\begin{abstract}
In contradiction to the simple AGN unification schemes, there exists a
significant population of broad line, 
$z\sim 2$ QSOs which have heavily absorbed
X-ray spectra. These objects have luminosities and redshifts characteristic of
the sources that produce the bulk of the QSO luminosity in the universe. Our
follow up observations in the submillimetre 
show that these QSOs are embedded in
ultraluminous starburst galaxies, unlike most unabsorbed QSOs at the same
redshifts and luminosities.  The radically different star formation properties
between the absorbed and unabsorbed QSOs implies that the X-ray absorption is
unrelated to the torus invoked in AGN unification schemes.
The most puzzling question about these objects is the nature of the X-ray
absorber.  We present our study of the X-ray absorbers based on deep 
(50--100ks)
XMM-Newton spectroscopy. The hypothesis of a normal QSO continuum, coupled with
a neutral absorber is strongly rejected. We consider the alternative
hypotheses for the absorber, originating either in the QSO or in the
surrounding starburst.
Finally we discuss the implications for QSO/host galaxy formation, in terms of
an evolutionary sequence of star formation and black hole growth. We propose
that both processes occur simultaneously in the gas-and-dust-rich heavily
obscured centres of young galaxies, and that absorbed QSOs form a transitional
stage, between the main obscured growth phase, and the luminous QSO.
\end{abstract}
\vspace{-2mm}

\section{Introduction}

\vspace{-3mm}
The prevalence of black holes in present day galaxy bulges, and the
proportionality between black hole and spheroid mass \citep{merritt01} implies
that the formation of the two components are intimately linked. One way to
probe star formation in distant QSOs is to observe them at submillimetre
wavelengths, and so measure the amount of radiation from young stars which is
absorbed and re-emitted by dust in the far infrared. With this in mind, we have
observed 
matched samples of X-ray absorbed and unabsorbed QSOs at 850$\mu$m with
SCUBA.  These observations revealed a remarkable dichotomy in the submillimetre
properties of these two groups of sources: X-ray absorbed QSOs are often
ultraluminous infrared galaxies, while X-ray unabsorbed QSOs are not. This
suggests that the two types are linked by an evolutionary sequence, whereby the
QSO emerges at the end of the main star-forming phase of a massive galaxy
\citep[][Carrera et~al. 2006, this volume]{page04,stevens05}. 

\vspace{-1mm}
However, the nature of the X-ray absorption remains
puzzling. It could be due to gas located within the AGN structure, or from more
distant material in the host galaxy. 
These objects are characterised by hard,
absorbed X-ray spectra, but they have optical/UV spectra which are typical for
QSOs, with broad emission lines and blue continua. Assuming that their 
hard X-ray
spectral shapes result from photoelectric absorption from cold material 
with solar
abundances, the column densities are $\sim 10^{22}$~cm$^{-2}$. These properties
are surprising: 
for a Galactic gas/dust ratio, the restframe ultraviolet spectra
would be heavily attenuated by such large columns of material. 
Therefore in order 
to investigate the X-ray absorption, we have obtained deep (50--100ks) {\em
  XMM-Newton} observations of three submillimetre bright, X-ray absorbed QSOs 
from our sample of hard-spectrum {\em
Rosat} sources \citep{page01a}. 
\vspace{-2mm}

\begin{figure}
\begin{center}
\leavevmode
\psfig{figure=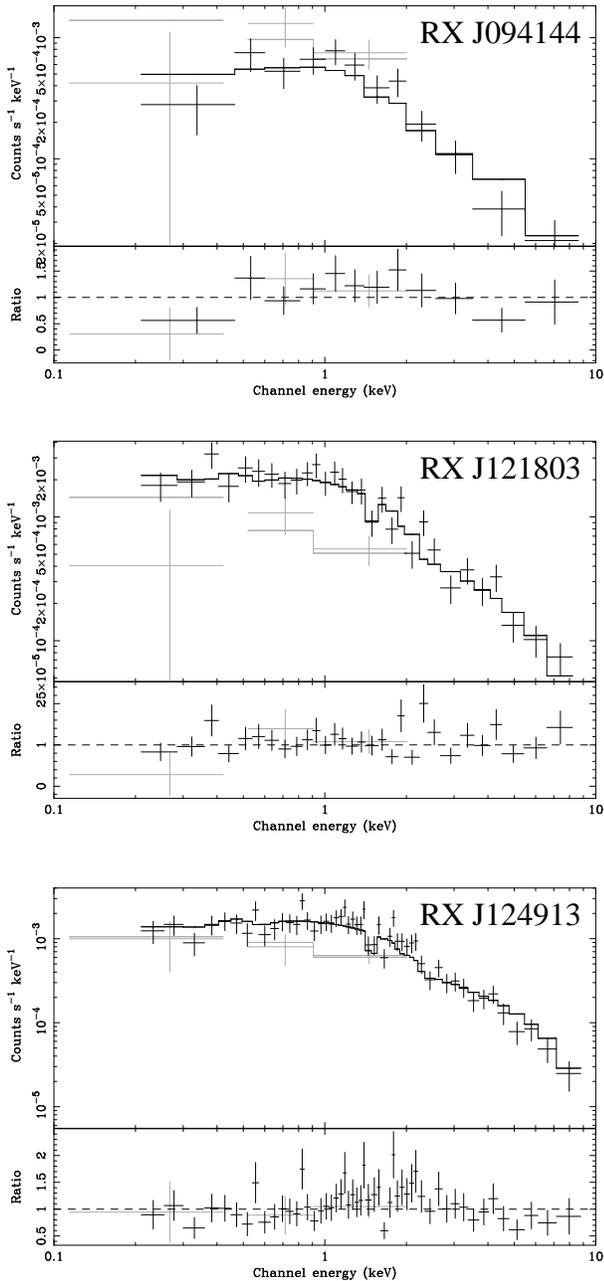,width=80truemm}
\vspace{3mm}
\caption{
{\em XMM-Newton} EPIC spectra (black) and {\em Rosat}
PSPC spectra (grey) of three X-ray absorbed QSOs.
The model is a simple power law with fixed Galactic absorption. 
The best-fit power law photon indices $\Gamma$ are $1.3\pm0.1$, $1.4\pm0.1$ and
$1.4\pm0.1$ for RX\,J094144, RX\,J121803, and RX\,J124913
respectively. Such photon indices are unusually hard for radio-quiet AGN. In all three objects there is a deficit of counts at the lowest
energy, indicating that absorption is responsible for the hard spectral shape;
this is also seen in the {\em Rosat} data. 
Furthermore, RX\,J094144 and RX\,J124913 show some systematic 
curvature relative to the power law model.
}
\label{fig:threeqsos}
 \end{center}
\end{figure}

\section{Results}

\vspace{-3mm}
The {\em XMM-Newton} spectra were first fitted with a power law and fixed
Galactic absorption. Surprisingly, the power law produces reasonable $\chi^{2}$
values. However, the photon indices are unusually hard for QSOs, and the data
show a deficit of counts relative to the model at the softest energies,
indicating that absorption is present. The original {\em Rosat} PSPC spectra
and the {\em XMM-Newton} spectra show excellent agreement 
(see Fig. \ref{fig:threeqsos}).

\vspace{-1mm}
The hypothesis of a normal ($\Gamma=2$) AGN X-ray spectrum and a cold absorber
is strongly rejected for RXJ121803 and RXJ124913. Therefore we considered
ionised absorber models for the X-ray absorption, using the `xabs' model in
{\small SPEX}, which includes both photoelectric and line absorption. For all
three AGN an acceptable fit can be obtained with a $\Gamma=2$ power law and an
absorber with an ionisation parameter log~$\xi \sim 2$ and column densities of
$10^{22.5}$--$10^{23.5}$~cm$^{-2}$. These absorbers have similar properties to
the high-ionisation absorber phases seen as outflows in some nearby Seyfert 1 
galaxies and
QSOs such as NGC3783 and PG1114+445 \citep{ashton04}.

\vspace{-1mm}
At these ionisation parameters and column densities, the absorbers are likely
to originate in the AGN themselves, rather than in the host galaxies. This
solution is attractive, because it is compatible with the lack of optical
extinction in these objects: if the absorber is driven as a wind, either from
the accretion disc or from evaporation of the inner edge of the molecular
torus, then dust will be sublimated before (or as) it enters the flow.
\vspace{-2mm}

\section{Implications for AGN and galaxy evolution}

\vspace{-3mm}
The low space density of X-ray absorbed QSOs relative to unabsorbed QSOs and to
distant ultraluminous galaxies detected in blank field SCUBA surveys, implies
that the X-ray absorbed QSOs are caught during a short-lived transitional
phase. Before this brief phase, AGN must be weak, and heavily obscured
\citep{alexander05}; after this phase the host galaxy is essentially fully
formed, and the naked QSO shines brightly until its fuel is consumed. A number
of theoretical models predict a very similar evolutionary pattern. In many of
these models, the QSO terminates the star formation in the host galaxy by
driving a powerful wind \citep[e.g.][]{fabian99,dimatteo05}. The EPIC
spectra of our X-ray absorbed QSOs suggest that the absorbers are
ionised winds driven by the AGN, and therefore that the transition between
buried AGN and naked QSO is mediated by a radiatively driven wind from the AGN,
as predicted by these models.
\vspace{-2mm}



%


\begin{thebibliography}{}
\bibitem[Alexander et~al.(2005)]{alexander05}
Alexander D.M., et~al., 2005, Nature, 433, 604 
\bibitem[Ashton et~al.(2004)]{ashton04}
Ashton C.E., et~al., 2004, MNRAS, 355, 73
\bibitem[Di Matteo Springel \& Hernquist(2005)]{dimatteo05}
Di Matteo T., Springel V. \& Hernquist L., 2005, Nature, 433, 604
\bibitem[Fabian(1999)]{fabian99}
Fabian A.C., 1999, MNRAS, 308, L39
\bibitem[Merritt \& Ferrarese(2001)]{merritt01}
Merritt D. \& Ferrarese L., 2001, MNRAS, 320, L30
\bibitem[Page, Mittaz \& Carrera(2001)]{page01a} 
Page M.J., Mittaz J.P.D. \& Carrera F.J., 
2001, MNRAS, 325, 575
\bibitem[Page et~al.(2004)]{page04}
Page M.J., Stevens J.A., Ivison R.J., Carrera F.J., 
2004, ApJ, 611, L85
\bibitem[Stevens et~al.(2005)]{stevens05}
Stevens J.A., et~al., 2005, MNRAS, 360, 610


\end{thebibliography}
\end{document}